\documentstyle[sprocl,psfig]{article}
\newenvironment{Eqnarray}%
       {\arraycolsep 0.14em\begin{eqnarray}}{\end{eqnarray}}
\newcommand{\beqa}{\begin{Eqnarray}}
\newcommand{\eeqa}{\end{Eqnarray}}
\newcommand{\beq}{\begin{equation}}
\newcommand{\eeq}{\end{equation}}

 \makeatletter

 \def\citenum#1{{\def\@cite##1##2{##1}\cite{#1}}}
 \def\@cite#1#2{[{#1}]}

 \newcount\@tempcntc
 \def\@citex[#1]#2{\if@filesw\immediate\write\@auxout{\string\citation{#2}}\fi
   \@tempcnta\z@\@tempcntb\m@ne\def\@citea{}\@cite{\@for\@citeb:=#2\do
     {\@ifundefined
        {b@\@citeb}{\@citeo\@tempcntb\m@ne\@citea\def\@citea{,}{\bf ?}\@warning
        {Citation `\@citeb' on page \thepage \space undefined}}%
     {\setbox\z@\hbox{\global\@tempcntc0\csname b@\@citeb\endcsname\relax}%
      \ifnum\@tempcntc=\z@ \@citeo\@tempcntb\m@ne
        \@citea\def\@citea{,}\hbox{\csname b@\@citeb\endcsname}%
      \else
       \advance\@tempcntb\@ne
       \ifnum\@tempcntb=\@tempcntc
       \else\advance\@tempcntb\m@ne\@citeo
       \@tempcnta\@tempcntc\@tempcntb\@tempcntc\fi\fi}}\@citeo}{#1}}
 \def\@citeo{\ifnum\@tempcnta>\@tempcntb\else\@citea\def\@citea{,}%
   \ifnum\@tempcnta=\@tempcntb\the\@tempcnta\else
    {\advance\@tempcnta\@ne\ifnum\@tempcnta=\@tempcntb \else
    \def\@citea{--}\fi
    \advance\@tempcnta\m@ne\the\@tempcnta\@citea\the\@tempcntb}\fi\fi}
\makeatother

\def\SCIPP{{\it Santa Cruz Institute for Particle Physics}\\
  {\it University of California, Santa Cruz, CA 95064 USA} \\}

\begin{document}
%
\thispagestyle{empty}
\begin{flushright}
{\large SCIPP 99/06}     \\[2pt]
{\large January, 1999}   \\[2pt]
{\large hep--ph/9901365}
\end{flushright}
\vskip1.5cm

\begin{center}
{\Large\bf  How~Well~Can~We~Predict~the~Mass~of~the~Higgs \\ [6pt] 
Boson of the Minimal Supersymmetric Model?}\\[1 cm]
{\large Howard E. Haber}\\[3pt]
{\it Santa Cruz Institute for Particle Physics  \\
   University of California, Santa Cruz, CA 95064, U.S.A.} \\[1.5cm]

{\bf Abstract}
\end{center}
\vskip0.25cm
The upper bound on the mass of the light CP-even Higgs boson of the 
minimal supersymmetric model (MSSM) depends on the supersymmetric 
particle spectrum via radiative loop effects.
At present, complete one-loop results and partial two-loop results
are known. Some simple analytic approximations have been obtained
which are quite accurate over a large portion of the MSSM
parameter space.  Based on these results, I examine how accurately
one can predict the upper bound on the mass of the lightest MSSM
Higgs boson.

\vskip3cm
\centerline{To appear in the Proceedings of the} 
\centerline{4th International Symposium on Radiative Corrections (RADCOR 98):}
\centerline{Applications of Quantum Field Theory to Phenomenology,} 
\centerline{Universitat Aut\'onoma de Barcelona,}
\centerline{Barcelona, Catalonia, Spain, 8-12 September 1998.}
\vfill
\clearpage
\setcounter{page}{1}

\title{
HOW WELL CAN WE PREDICT THE MASS OF THE HIGGS BOSON OF THE MINIMAL
SUPERSYMMETRIC MODEL? }
\author{HOWARD E. HABER}
\address{\SCIPP}

\maketitle\abstracts{The upper bound on the mass of the light
CP-even Higgs boson of the minimal supersymmetric model (MSSM)
depends on the supersymmetric particle spectrum via radiative loop
effects.
At present, complete one-loop results and partial two-loop results
are known. Some simple analytic approximations have been obtained
which are quite accurate over a large portion of the MSSM
parameter space.  Based on these results, I examine how accurately
one can predict the upper bound on the mass of the lightest MSSM
Higgs boson. }
\def\lsim{\mathrel{\raise.3ex\hbox{$<$\kern-.75em\lower1ex\hbox{$\sim$}}}}
\def\gsim{\mathrel{\raise.3ex\hbox{$>$\kern-.75em\lower1ex\hbox{$\sim$}}}}
\def\ifmath#1{\relax\ifmmode #1\else $#1$\fi}
\def\half{\ifmath{{\textstyle{1 \over 2}}}}
\def\quarter{\ifmath{{\textstyle{1 \over 4}}}}
\def\3quarter{{\textstyle{3 \over 4}}}
\def\ninequarters{{\textstyle{9 \over 4}}}
\def\ninehalves{{\textstyle{9 \over 2}}}
\def\threehalves{{\textstyle{3 \over 2}}}
\def\threequarters{{\textstyle{3 \over 4}}}
\def\threeeighths{{\textstyle{3 \over 8}}}
\def\third{\ifmath{{\textstyle{1 \over 3}}}}
\def\eighth{\ifmath{{\textstyle{1 \over 8}}}}
\def\eightthirds{\ifmath{{\textstyle{8 \over 3}}}}
\def\twothirds{{\textstyle{2 \over 3}}}
\def\fourthirds{{\textstyle{4 \over 3}}}
\def\fortythirds{{\textstyle{40 \over 3}}}
\def\fivesixths{\ifmath{{\textstyle{5\over 6}}}}
\def\fourth{\ifmath{{\textstyle{1\over 4}}}}
\def\sqhalf{\ifmath{{\textstyle{1 \over \sqrt{2}}}}}
\def\tanb{\tan\beta}
\def\calo{{\cal{O}}}
\def\calm{{\cal{M}}}
\def\calmm{{\calm}^2}
\def\hl{h^0}
\def\ha{A^0}
\def\hh{H^0}
\def\hpm{H^\pm}
\def\mha{m_{\ha}}
\def\mhl{m_{\hl}}
\def\mhh{m_{\hh}}
\def\mhpm{m_{\hpm}}
\def\mz{m_Z}
\def\mw{m_W}
\def\mt{m_t}
\def\mb{m_b}
\def\msusy{M_{\rm SUSY}}
\def\msusyy{M_{\rm SUSY}^2}
\def\tr{\rm tr}
\def\cw{\cos\theta_W}
\def\sw{\sin\theta_W}
\def\swiv{\sin^4\theta_W}
\def\ctwob{\cos2\beta}
\def\ctwobb{\cos^2 2\beta}
\def\d1l{\delta\lambda_1}
\def\d2l{\delta\lambda_2}
\def\d3l{\delta\lambda_3}
\def\d4l{\delta\lambda_4}
\def\d5l{\delta\lambda_5}
\def\d6l{\delta\lambda_6}
\def\d7l{\delta\lambda_7}
\def\sb{\sin\beta}
\def\cb{\cos\beta}
\def\sbb{\sin^2\beta}
\def\cbb{\cos^2\beta}
\def\sbiv{\sin^4\beta}
\def\cbiv{\cos^4\beta}
\def\mzz{m_Z^2}
\def\mww{m_W^2}
\def\sw#1{\sin^{#1}\theta_W}
\def\cw#1{\cos^{#1}\theta_W}
\def\refs#1#2{refs.~\cite{#1} and \cite{#2}}
\def\Ref#1{ref.~\cite{#1}}
\def\Rref#1{Ref.~\cite{#1}}
\def\eq#1{eq.~(\ref{#1})}
\def\Eq#1{Eq.~(\ref{#1})}
\def\eqs#1#2{eqs.~(\ref{#1}) and (\ref{#2})}
\def\eqss#1#2#3{eqs.~(\ref{#1}), (\ref{#2}) and (\ref{#3})}

%
%
%
\section{Introduction}

Low-energy supersymmetry \cite{susy}
provides the most compelling framework for electroweak physics,
in which the electroweak symmetry breaking is generated via the
dynamics of an elementary scalar Higgs sector.  The scalar boson
masses are kept light (of order the electroweak symmetry breaking
scale) due to an approximate supersymmetry in nature.
Supersymmetry is broken at the TeV scale or below, and this
information is transmitted to the scalar sector, thereby
generating electroweak symmetry breaking dynamics at the proper scale.

The simplest model of low-energy supersymmetry is the minimal
supersymmetric extension of the Standard Model (MSSM). In this
model, the Higgs sector consists of eight degrees of freedom
made up from two complex weak scalar doublets of hypercharge $\pm
1$ respectively \cite{hhg}. Supersymmetry requires that the hypercharge
$-$1 [+1] Higgs doublets couple exclusively to down-type [up-type]
fermions,
respectively.  After minimizing the Higgs potential, the neutral
components of the Higgs doublets acquire vacuum expectation values
(vevs) with $\langle H_i\rangle=v_i/\sqrt{2}$.  
The model possesses five physical Higgs bosons: two CP-even scalars,
$\hl$ and $\hh$ (with $\mhl<\mhh$), a CP-odd Higgs scalar $\ha$
and a charged Higgs pair $H^\pm$. As usual, I define
$\tan\beta\equiv v_2/v_1$ and normalize $v^2\equiv
v_1^2+v_2^2\equiv 4m_W^2/g^2=(246~{\rm GeV})^2$. Due to the form
of the Higgs-fermion interaction, the third
generation quark masses are given by $m_b= h_b v_1/\sqrt{2}$ and
$m_t= h_t v_2/\sqrt{2}$, where $h_q$ ($q=t,b$) are the
corresponding Yukawa couplings.

The tree-level physical Higgs spectrum is easily computed
\cite{hhg}.  Its
most noteworthy feature is the upper bound on the light CP-even
Higgs scalar: $\mhl\leq\mz|\cos 2\beta|\leq\mz$.  The maximum
tree-level upper bound of $\mz$ is saturated when one of the vevs
vanishes (and $\mha>\mz$).
It is convenient to consider a limiting case  of the
MSSM Higgs sector where $v_1=0$.  For finite $h_b$, this
limit corresponds to $m_b=0$, which is a reasonable
approximation.\footnote{In practice, it is sufficient to take
$v_1\ll v_2$, and then fix the value of $h_b$ to be consistent
with the observed $b$-quark mass.} In the $v_1=0$ model, the Higgs
sector degenerates to a one-doublet model with: 
\beq
\label{vonedoublet} V_{\rm
Higgs}=m^2\Phi^\dagger\Phi+\half\lambda(\Phi^\dagger\Phi)^2\,,
\qquad\lambda\equiv\quarter(g^2+g^{\prime 2})\,. 
\eeq 
The supersymmetric constraint on the value of $\lambda$ is a
consequence of the fact that the MSSM Higgs quartic couplings
originate from the $D$-term contributions to the scalar potential.
The squared-mass of the light CP-even Higgs boson of the $v_1=0$ model is
given by $\mhl^2=\lambda v^2=\mz^2$.

\section{Upper bound of {\boldmath $\mhl$} in the MSSM}

The upper bound of $\mhl\leq\mz$ will be modified by radiative
corrections \cite{hhprl,early-veff}.  In order to obtain the
radiatively corrected
upper bound of $\mhl$, it suffices to compute radiative corrections in
the $v_1=0$ model.  Let us focus on the real part of the neutral
scalar component: $\Phi^0=(v+h)/\sqrt{2}$.
The bare Higgs potential takes the following form: $V_{\rm
Higgs}=t_0 h+\half (\mhl^2)_0 h^2 + {\cal O}(h^3)$, where
\beq\label{bareparms}
t_0 = v_0\left[\half\lambda_0
v_0^2+m_0^2\right]\,,\qquad (\mhl^2)_0 =
m_0^2+\threehalves\lambda_0 v_0^2\,,
\eeq
and the subscript $0$
indicates bare parameters. We also introduce 
\beq
(\mz^2)_0=\quarter (g_0^2+g_0^{\prime 2})v_0^2\,.
\eeq
The
on-shell renormalization scheme is defined such that $\mz$ and
$\mhl$ are physical masses corresponding to zeros of the
corresponding inverse propagators. Let the sum of all one-loop
(and higher) Feynman graphs contributing to the $Z$ and $\hl$
two-point functions be denoted by
$iA_{ZZ}(q^2)g^{\mu\nu}+iB_{ZZ}(q^2)q^\mu q^\nu$ and
$-iA_{hh}(q^2)$, respectively, where $q$ is the four-momentum of
one of the external legs.  The physical masses are given by:
\beqa
\mz^2 &=& (\mz^2)_0+{\rm Re}~A_{ZZ}(\mz^2)\,, \label{oneloopz}\\
\mhl^2 &=& (\mhl^2)_0+{\rm Re}~A_{hh}(\mhl^2)\,.\label{onelooph}
\eeqa
Since $v$ is the vev
of the scalar field at the true minimum of the potential, we
require that the sum of all tadpoles must vanish.  That is,
\beq\label{tadpole}
t_0+A_h(0)=0\,,
\eeq
where $-iA_h(0)$ is the sum of
all one-loop (and higher) Feynman graphs contributing to the $\hl$
one-point function.  Combining
eqs.~(\ref{bareparms})--(\ref{tadpole}), we obtain
\beq\label{oneloopmass}
\mhl^2=\mz^2+{\rm
Re}~\left[A_{hh}(\mz^2)-A_{ZZ}(\mz^2)\right] -{A_h(0)\over v}
+\left[\lambda_0-\quarter(g_0^2+g_0^{\prime 2})\right]v_0^2\,.
\eeq
This result is accurate at one-loop order, since we have put
$\mhl=\mz$ and $v_0=v$ on the right hand side where possible.

Naively, one might argue that \eq{oneloopmass} can be
simplified by using the supersymmetric condition
$\lambda_0=\quarter(g_0^2+g_0^{\prime 2})$.  However,
this is correct only if a regularization scheme that preserves
supersymmetry is employed.  Of course, the physical quantity
$\mhl^2$ must be independent of scheme.
Consider two different regularization schemes:
dimensional regularization (DREG) and
dimensional reduction \cite{dred} (DRED).  Renormalized couplings
defined via (modified) minimal subtraction in these two schemes are
called $\overline{\rm MS}$ and $\overline{\rm DR}$ couplings,
respectively.  DREG does not preserve supersymmetry because the number
of gauge and gaugino degrees of freedom does not match in $n\neq 4$
dimensions.  In contrast, DRED preserves supersymmetry
(at least at one and two-loop order).
In DRED [DREG], bare quantities will be denoted
with the subscript $D$ [$G$].  Then,
the supersymmetric condition holds in DRED:
\beq \label{dred}
\lambda_D-\fourth(g_{D}^2+g_{D}^{\prime 2})=0 \,.
\eeq

We now demonstrate that the above relation is
violated in DREG.  First, the gauge couplings of the two schemes are
related as follows \cite{antoniadis}
\beq \label{gdg}
g_{D}^2 = g_{G}^2+{g^4\over 24\pi^2}\,,\qquad\qquad
g_{D}^{\prime 2}  =  g_{G}^{\prime 2}\,.
\eeq
For the Higgs self-coupling $\lambda$,
the relation between the two schemes is derived by considering
the one-loop effective potential (in the Landau gauge),
$V\equiv V^{(0)}+V^{(1)}$, where $V^{(0)}$ is the
tree-level scalar potential and $V^{(1)}$ is given by:
\beq \label{vone}
V^{(1)} =-{1\over 64 \pi^2}\, {\rm Str}\,\calm^4(\Phi)\left[\Delta +K - \ln
{\calm^2(\Phi)\over \mu^2}\right]\,.
\eeq
In \eq{vone}, $K$ is a scheme-dependent constant (see below),
$\calm^2(\Phi)$ denotes the squared-mass
matrix as a function of the scalar Higgs fields ({\it i.e.},
the corresponding tree-level squared-mass matrices are obtained when
$\Phi$ is replaced by its vev),
and
the divergences that arise in the computation of the one-loop integrals
in $4-2\epsilon$ dimensions appear in the factor
$\Delta\equiv 1/\epsilon
-\gamma_E+\ln(4\pi)$ [where $\gamma_{\rm E}$ is the Euler constant].
We have also employed the notation
\beq \label{strace}
{\rm Str}\,\{\cdot\cdot\cdot\}\equiv\sum_i\,C_i(2J_i+1)(-1)^{2J+1}\,
\{\cdot\cdot\cdot\}_i\,,
\eeq
where the sum is taken over the corresponding mass matrix eigenvalues,
including a factor $C_i$ which counts internal degrees of freedom
({\it e.g.}, charge and color) for all particles of spin $J_i$ that
couple to the Higgs bosons.

In DRED, $K=3/2$, independent of particle $i$ in the sum [\eq{strace}].
The fact that particles of different spin yield the same constant $K$ is an
indication that DRED preserves supersymmetry at one-loop.
In DREG, $K=3/2$ for spin 0 and spin-1/2 particles, while
$K=5/6$ for spin-1 particles.  However, the effective
potential (expressed in terms of bare parameters) must be independent of
scheme.  Comparing the DREG and DRED computations, it follows that
\beq
\eighth\lambda_D v^4 -{1\over 64\pi^2}
\left(\threehalves\right)(6m_W^4+3m_Z^4)
=\eighth\lambda_G v^4-{1\over 64\pi^2}
\left(\fivesixths\right)(6m_W^4+3m_Z^4)\,,
\eeq
which yields
\beq \label{lambdadg}
\lambda_D=\lambda_G+{g^4(\mz^4+2m_W^4)\over 64\pi^2m_W^4}\,.
\eeq
Combining the results of \eqs{gdg}{lambdadg} gives the DREG result
\beq \label{lambdarel}
\lambda_G-\quarter(g_{G}^2+g_{G}^{\prime 2})=-{g^4\over 64\pi^2m_W^4}
\left(\mz^4 +\fourthirds m_W^4\right)\,.
\eeq
Thus, in computing the one-loop corrected Higgs mass [\eq{oneloopmass}]
in DREG [DRED], one must use the relation between $\lambda_0$ and
$\quarter(g_0^2+g_0^{\prime 2})$ given be \eq{lambdarel} [\eq{dred}].
One can check that this difference is precisely compensated by the
difference in DREG and DRED that arises in
the computation of the vector boson loop contributions to
the one-point and two-point functions.
Henceforth, we shall always use the DRED scheme, in which case
\cite{hhprl}
\beq \label{dredoneloopmass}
\mhl^2=\mz^2+{\rm Re}~\left[A_{hh}(\mz^2)-A_{ZZ}(\mz^2)\right]
-{A_h(0)\over v}\,.
\eeq
Although the loop functions above are individually divergent,
all divergences precisely cancel in the sum and yield a well-defined
one-loop result for $\mhl$.

The method described above [resulting in \eq{dredoneloopmass}] is
sometimes called the diagrammatic method \cite{hhprl,completeoneloop}
since one explicitly evaluates
the one-point and two-point functions by standard Feynman diagram
techniques.  A second method for computing $\mhl$, called the effective
potential technique,
is often employed in the literature \cite{early-veff,veff,carena}.  This
is not an alternate ``scheme'', but simply another way of organizing the
calculation.  Consider the DRED one-loop effective potential introduced
above (with $K=3/2$).  In the sum $V=V^{(0)}+V^{(1)}$, the
$\overline{\rm DR}$ scheme consists of absorbing the factor of $\Delta$ into
the bare parameters ($m_0$, $\lambda_0$ and $\Phi_0$),
which converts them into $\overline{\rm DR}$ parameters.  
Renormalized quantities (such as the effective potential or the
$n$-point Green functions) will be denoted with tildes in the following.
These are computed in the Landau gauge; the divergent piece $\Delta$
is removed by $\overline{\rm DR}$ subtraction and the bare parameters
are replaced by renormalized $\overline{\rm DR}$ parameters.  Finally,
the $\overline{\rm DR}$ parameters are related to physical
parameters.  

We proceed as follows.
First, we minimize the renormalized
effective potential by setting the first derivative equal to zero.  This
condition yields:
\beq \label{mincon}
\left[\half\lambda v^2+m^2\right]v+\widetilde A_h(0)=0\,.
\eeq
In \eq{mincon}, the first term on the left hand side arises at
tree-level, while the second term is a consequence of the
the fact that the $n$th derivative of
$V^{(1)}$, evaluated at the potential minimum, is equal to the scalar
$n$-point function evaluated at zero external momentum \cite{pokorski}.
The second
derivative of the effective potential, denoted by $(\mhl^2)_{\rm eff}$,
is similarly given by:
\beq \label{meff}
(\mhl^2)_{\rm eff}=m^2+\threehalves\lambda v^2+\widetilde A_{hh}(0)\,.
\eeq
We may use the DRED relation [\eq{dred}], which is also satisfied by the
renormalized $\overline{\rm DR}$ parameters, to eliminate
$\lambda$.  The $\overline{\rm DR}$ $Z$-mass parameter is given by
$(\mz^2)_{\overline{\rm DR}}=\quarter (g^2+g^{\prime 2})v^2$.  Combining
the above results yields:
\beq \label{mhone}
(\mhl^2)_{\rm eff}= (\mz^2)_{\overline{\rm DR}}+\widetilde A_{hh}(0)-
{\widetilde A_h(0)\over v}\,.
\eeq

In the literature, $(\mhl^2)_{\rm eff}$ is sometimes used as the
approximation to the one-loop-improved Higgs squared-mass.
However, this is {\it not} a physical parameter, since it 
depends on an arbitrary scale that is introduced in
the $\overline{\rm DR}$ subtraction scheme.  To obtain an expression
for the physical mass, which corresponds to the zero of inverse propagator,
we note that $(\mhl^2)_{\rm eff}$ has been computed using the two-point
function evaluated at {\it zero} external momentum.
Thus, the physical Higgs squared-mass is given by:
\beq \label{mhtwo}
\mhl^2=(\mhl^2)_{\rm eff}+{\rm Re}~\widetilde A_{hh}(\mhl^2)-\widetilde
A_{hh}(0)\,.
\eeq
Likewise, we must convert from $(\mz^2)_{\overline{\rm DR}}$ to the
physical $Z$ squared-mass.  This is accomplished using a result
analogous to that of \eq{oneloopz}, which guarantees that $m_Z$ corresponds
to the zero of the inverse $Z$ propagator:
\beq \label{mhthree}
\mz^2=(\mz^2)_{\overline{\rm DR}}+{\rm Re}~\widetilde A_{ZZ}(\mz^2)\,.
\eeq
Combining eqs.~(\ref{mhone})--(\ref{mhthree}), we end up with
\beq \label{veffoneloopmass}
\mhl^2=\mz^2+{\rm Re}~\left[\widetilde A_{hh}(\mz^2)-\widetilde A_{ZZ}(\mz^2)\right]
-{\widetilde A_h(0)\over v}\,.
\eeq
Not surprisingly, we have reproduced the diagrammatic result
[\eq{dredoneloopmass}].


\section{Leading Logarithms and Renormalization Group
Improvement}

When the loop functions in \eq{dredoneloopmass} are computed, one finds
that the most significant contributions grow logarithmically with the
top squark masses.  (Terms that are logarithmically sensitive to other
supersymmetric particle masses also exist.)   Over a large range
of supersymmetric parameter space, the
radiatively corrected Higgs mass can be
well approximated by just a few terms.  On the other hand, if the
logarithms become too large, then the
validity of the perturbation theory becomes suspect.  However, in this
case the leading
logarithms can be resummed using renormalization group (RG)
techniques \cite{rgesum,llog}.

We begin with a one-loop analysis.  Consider an effective field theory
approach \cite{llog}, and assume for simplicity that supersymmetry
breaking
is characterized by one mass scale, $\msusy$, which is assumed to be
large compared with $\mz$.  At scales $\mu\leq\msusy$,
the Higgs potential takes the form:
\beq
V=\half m^2(\mu)[h(\mu)]^2+\eighth\lambda(\mu)[h(\mu)]^4\,.
\eeq
Letting $h\to h+v$ with $m^2(\mu)<0$, the Higgs mass is given by
\beq \label{mhlmu}
\mhl^2(\mu)=\lambda(\mu) v^2(\mu)\,.
\eeq
Since the effective theory is supersymmetric only for $\mu\geq\msusy$,
we impose the supersymmetric boundary condition [see \eq{dred}]:
\beq \label{boundary}
\lambda(\msusy)=\quarter\left[g^2(\msusy)+g^{\prime 2}(\msusy)\right]\,.
\eeq
Scale dependent parameters satisfy renormalization group equations
(RGEs).  For $\mu<\msusy$, the Standard Model RGEs are relevant:
\beqa \label{rges}
\beta_\lambda &\equiv& {d\lambda\over d\ln \mu^2} = {1\over
16\pi^2}\Bigl[6\lambda^2+\threeeighths[2g^4+(g^2+g^{\prime 2})^2]-2\sum_f
N_{cf} h_f^4\Bigr] -2\lambda\gamma_v\,, \nonumber \\
\gamma_v &\equiv& {d\ln v^2\over d\ln \mu^2}
 =  {1 \over 16\pi^2}\Bigl[\ninequarters g^2 + \threequarters
g^{\prime 2}-\sum_f N_{cf} h_f^2\Bigr]\,, \nonumber \\
\beta_{g^2+g^{\prime 2}} & \equiv & {d(g^2+g^{\prime 2}) \over d\ln \mu^2}
= {1\over 96\pi^2}\Bigl[(8N_g-43)g^4
+(\fortythirds N_g+1)g^{\prime 4}\Bigr]\,,
\eeqa
where $h_f=\sqrt{2} m_f/v$, $N_g=3$ is the number of fermion
generations, and $N_{cf}=3$ $[1]$ when $f$ runs over quark [lepton] 
indices.

It is instructive to solve the RGEs iteratively to one-loop, by ignoring the 
$\mu$ dependence on the right hand sides in \eq{rges}.  
Incorporating the boundary condition [\eq{boundary}],
the solution for $\lambda(\mz)$ is given by~\footnote{One subtlety
consists of the proper way to run down from $m_t$ to $m_Z$, since
below $\mu=\mt$, the electroweak symmetry is broken.  I will ignore 
this subtlety here although it can be addressed; see \Ref{llog}.}
\beqa
\lambda(\mz) & = & \quarter(g^2+g^{\prime
2})(\msusy)-\beta_\lambda\ln\left({\msusyy\over\mz^2}\right) \nonumber \\
& = & \quarter(g^2+g^{\prime
2})(\mz)+(\quarter\beta_{g^2+g^{\prime 2}}
-\beta_\lambda)\ln\left({\msusyy\over\mz^2}\right)\,.
\eeqa

Finally, using \eq{mhlmu}, we identify the physical Higgs mass by
evaluating $\mhl^2(\mu)$ at $\mu=\mz$ and taking $v(\mz)= 246$~GeV.
We know from the previous section that this is not strictly correct.
However, at the one-loop leading logarithmic level, this procedure is
accurate, and we end up with:
\beq \label{mh1LL}
(\mhl^2)_{\rm 1LL}= \mz^2+ (\quarter\beta_{g^2+g^{\prime 2}}
-\beta_\lambda)v^2\ln\left({\msusyy\over\mz^2}\right)\,,
\eeq
where the subscript 1LL indicates that the result is only accurate to
one-loop leading logarithmic order.  To obtain the full one-loop leading
logarithmic expression, simply insert the results of \eq{rges} into
\eq{mh1LL} [in $\beta_\lambda$ one can consistently
set $\lambda=\quarter(g^2+g^{\prime 2})$].
We have checked \cite{hhprl,llog}
that the above result matches precisely with
the diagrammatic computation [\eq{dredoneloopmass}] in the limit of
$\msusy\gg\mz$, where $\msusy$ characterizes the scale of 
supersymmetric particle masses (taken to be roughly degenerate).
The dominant term at one-loop is proportional to $m_t^4$ and
arises from the piece of $\beta_\lambda$ proportional to $h_t^4$.
Inserting $\beta_\lambda=-3h_t^4/8\pi^2$ with
$h_t=\sqrt{2}m_t/v$ into \eq{mh1LL}, one obtains~\footnote{The lower
scale of the logarithm in this case is $m_t^2$ (and not $\mz^2$) since
this term arises from the incomplete cancelation of the top quark and
top squark loops.}
\beq \label{mhlapprox}
(\mhl^2)_{\rm 1LT}=\mz^2+{3g^2\mt^4\over
8\pi^2\mw^2}\ln\left(\msusy^2\over\mt^2\right)\,.
\eeq
The subscript 1LT indicates that this is the leading $m_t^4$ piece of
$(\mhl^2)_{\rm 1LL}$.  However, the additional terms in $(\mhl^2)_{\rm
1LL}$ are numerically significant as we shall show at the end of this
section.

Thus, we see that given the RG functions, no additional diagrammatic
computations are needed to extract the full one-loop leading logarithmic
contribution to the Higgs mass.  Thus the RG-approach provides a useful
short cut for extracting the leading one-loop contributions to the Higgs
mass.  Of course, if the leading logarithms are
large, then they should be resummed to all orders.  
This is accomplished by computing the
RG-improvement of the exact one-loop result as follows.  Let
$(\mhl^2)_{\rm 1RG}\equiv\lambda(\mz)v^2(\mz)$,
where $\lambda(\mz)$ is obtained by {\it numerically} 
solving the one-loop RGEs.  Write the exact
one-loop result as: $\mhl^2=(\mhl^2)_{\rm 1LL}+(\mhl^2)_{\rm 1NL}$,
where $(\mhl^2)_{\rm 1NL}$ is the result obtained by subtracting the
one-loop leading logarithmic contribution from the exact one-loop
result.  Clearly, this piece contains no term that grows logarithmically with
$\msusy$.  Then the complete one-loop RG-improved result is given by
$\mhl^2=(\mhl^2)_{\rm 1RG}+(\mhl^2)_{\rm 1NL}$.

The RG technique can be extended to two loops as follows \cite{hhh}.
For simplicity, we focus on the leading corrections, which depend on
$\alpha_t\equiv h_t^2/4\pi$ and $\alpha_s\equiv g_s^2/4\pi$, {\it
i.e.}, we work in the approximation of
$h_b=g=g'=0$ and $\lambda\ll h_t$.  (All two-loop results quoted in this
section are based on this approximation.)
The dependence on the strong coupling constant is a new
feature of the two-loop analysis.  
We now solve the one-loop RGEs by iterating twice to two loops.
In the second iteration, we need the RGE for $h_t^2$, which in the
above approximation is given by
\beq \label{rgeht}
\beta_{h_t^2}\equiv\frac{\rm d}{\rm d\ln\mu^2}\,h_t^2
 = \frac{1}{16\pi^2}\left[\ninehalves h_t^2 -8 g_s^2\right]\,h_t^2\,.
\eeq
This iteration produces the two-loop leading double logarithm
\cite{carena}, and yields
\begin{equation} \label{lambdatwice}
\lambda(\mt)=\quarter(g^2+g^{\prime 2})-{3h_t^4(\mt)\over 8\pi^2}\ln\left(
{\msusyy\over\mt^2}\right)\left[1+\left(\gamma_v+
{\beta_{h_t^2}\over h_t^2}\right)\ln\left({\msusyy\over\mt^2}\right)
\right]\,.
\end{equation}

Next, we must incorporate the sub-dominant two loop effects.
Only three modifications of our one-loop analysis
are required (in the limit of $h_b=g=g'=0$ and $\lambda\ll h_t$).
First, we need only the $h_t$ and $g_s$ dependent parts of the
two loop contribution to
$\beta_\lambda$.  That is, $\beta_\lambda$ is
modified as follows \cite{2looprges}
\begin{equation}
\beta_{\lambda} \longrightarrow \beta_{\lambda} +
  \frac{1}{(16\pi^2)^2}\,\left[\,30 h_t^6-32 h_t^4\,g_s^2\,\right]\,.
\label{betalambdatwoloop}
\end{equation}
Including this into the iterative solution of the RGEs adds a two-loop
singly logarithmic term to the result of \eq{lambdatwice}.
Second, we must
distinguish between the Higgs pole mass (denoted by $\mhl$ with
no argument) and the running Higgs mass evaluated at $\mt$.
Using the results of Sirlin and Zucchini \cite{sirlin},
\begin{equation}
\mhl^2  =  {4\,\mw^2\,\lambda(\mt)\over g^2}\left[1+\frac{1}{8}
 \left(\frac{\alpha_t}{\pi}\right)\right]\,.
\label{poletorun}
\end{equation}
Third, we make use of the relation
between $v^2(\mt)$ and $v^2\equiv 4\mw^2/g^2$,
\begin{equation} \label{vevcorrection}
v^2(\mt)={4\mw^2\over g^2}\left[1-{3\over 8}\left({\alpha_t\over\pi}\right)
\right]\,.
\end{equation}
Using the above results, we end up with
\beqa \label{twolooprun}
\mhl^2 & =& \mz^2+
  \frac{3g^2}{8\pi^2\mw^2}\,
  m_t^4(m_t)\,
    \ln\left(\frac{\msusyy}{m_t^2}\right)\,
  \left[1+\left(\gamma_v+
        \frac{\beta_{h_t^2}}{h_t^2}\right)\,
        \ln\left(\frac{\msusyy}{m_t^2}\right)\right. \nonumber \\
      &&\qquad\qquad\qquad\qquad\qquad +
        \left.\frac{4}{3}\left(\frac{\alpha_s}{\pi}\right)-
        \frac{3}{8}\left(\frac{\alpha_t}{\pi}\right)\,\right] \,,
\eeqa
where $h_t\equiv h_t(m_t)$ and $m_t(m_t)\equiv h_t(m_t)\,v(m_t)/\sqrt{2}$.
Numerically, the two-loop singly logarithmic piece of \eq{twolooprun}
contributes about $3\%$ relative to the one-loop leading logarithmic
contribution.

Let us compare this result with the
two-loop diagrammatic computation of \Ref{hempfhoang}.  In order to
make this comparison, we must express \eq{twolooprun} in terms of the
top quark pole mass, $m_t$.  The relation between $m_t$ and the running
top-quark mass is given by
\cite{tpoleref1,tpoleref2}
\begin{equation} \label{tpole}
m_t =  m_t(m_t)\,\left[1+
    \frac{4}{3}\left(\frac{\alpha_s}{\pi}\right)-
    \frac{1}{2}\left(\frac{\alpha_t}{\pi}\right)\right]\,,
\end{equation}
where $m_t(m_t)$ is the $\overline{\rm MS}$ running top-quark mass
evaluated at $m_t$.\footnote{We
caution the reader that \Ref{tpoleref2} defines
$m_t(m_t)=h_t(m_t)v/\sqrt{2}$, which differs slightly from
the definition of $m_t(m_t)$ used here.}
Inserting the above result into \eq{twolooprun} yields:
\beqa \label{leading2loop}
\mhl^2 & =& \mz^2+
  \frac{3g^2 m_t^4}{8\pi^2\mw^2}\,
  \ln\left(\frac{\msusyy}{m_t^2}\right)\,
  \left[1+\left(\gamma_v+
        \frac{\beta_{h_t^2}}{h_t^2}\right)\,
        \ln\left(\frac{\msusyy}{m_t^2}\right) \right. \nonumber \\
   &&     \qquad\qquad\qquad\qquad\qquad  -\left.
        \left(\frac{4\alpha_s}{\pi}\right)+
        \frac{13}{8}\left(\frac{\alpha_t}{\pi}\right)\,\right] \,.
\eeqa
This result matches precisely the one obtained in \Ref{hempfhoang} in the
limit of $\msusy\gg\mz$.  Note that the numerical
contribution of the two-loop singly-logarithmic contribution in
\eq{leading2loop} is about $10\%$ of the corresponding one-loop
contribution. Clearly, the
use of the running top quark mass [as in \eq{twolooprun}]
results in a slightly
better behaved perturbation expansion.

Finally, we can employ a very useful trick to make our results above
even more compact.  The two-loop doubly-logarithmic contribution can be
absorbed into the one-loop leading-logarithmic contribution by an
appropriate choice of scale for the running top-quark mass.
Specifically, using the iterative one-loop leading-logarithmic
solution to the RGEs for $h_t$ and $v$ yields
\beq
m_t(\mu)  =  \sqhalf h_t(\mu) v(\mu) =
m_t(m_t)\left[1-\left({\alpha_s\over\pi}-{3\alpha_t\over
16\pi}\right)\ln\left({\mu^2\over m_t^2}\right)\right]\,.
\eeq
If we choose the scale $\mu_t\equiv\sqrt{m_t\msusy}$ to evaluate the
running top-quark mass in \eq{twolooprun}, we end up with:
\begin{equation}
\mhl^2  =  \mz^2+
  \frac{3g^2}{8\pi^2\mw^2}\,
  m_t^4(\mu_t)\, \ln\left(\frac{\msusyy}{m_t^2(\mu_t)}\right)\,
  \left[1+
        \frac{1}{3}\left(\frac{\alpha_s}{\pi}\right)-
        \frac{3}{16}\left(\frac{\alpha_t}{\pi}\right)\,\right]\,.
\label{higgsmass3}
\end{equation}
One can check that the sum of the terms in the brackets
deviates from one by less than $1\%$.  Thus, in practice, the
two-loop singly-logarithmic contribution can now be neglected since it
is numerically insignificant.  That is, one can incorporate the
leading two-loop contributions by
simply inserting the running top-quark mass evaluated
at $\mu_t\equiv\sqrt{m_t\msusy}$ into
the one-loop leading-logarithmic expression for $\mhl^2$.

\begin{figure}[htb]
\centerline{\psfig{file=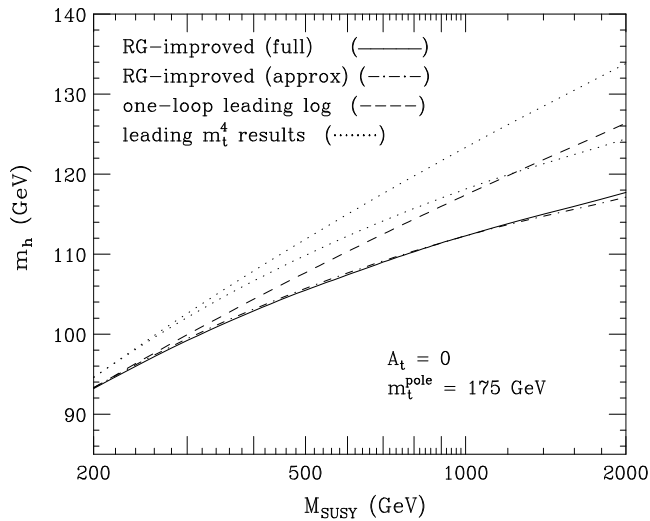,width=8.5cm,height=6.8cm,angle=90}}
\caption{The upper bound to the mass of the light CP-even Higgs boson
of the MSSM is plotted as a function of the common supersymmetric mass
$\msusy$ (in the absence of squark mixing).
The one-loop leading logarithmic result [dashed line]
is compared with the RG-improved result, which was obtained
by a numerical computation [solid line] and by the simple
recipe described in the text [dot-dashed line].
Also shown are the leading $m_t^4$ result of
eq.~(\protect\ref{mhlapprox}) [higher dotted line], and its
RG-improvement [lower dotted line].  The running top quark mass used in
our numerical computations is $m_t(m_t)= 166.5$~GeV.}
\label{hhhfig1}
\end{figure}

Fig.~\ref{hhhfig1} illustrates the results of this section.  We
display the results for $\mhl$ based on
five different expressions for the light CP-even Higgs mass.
Case~(i) corresponds to the one-loop leading $m_t^4$ result,
$(\mhl^2)_{\rm 1LT}$ [\eq{mhlapprox}].  In case~(ii) 
we exhibit the full one-loop leading logarithmic expression,
$(\mhl^2)_{\rm 1LL}$ [\eq{mh1LL}]. In case~(iii), we consider
$(\mhl^2)_{\rm 1RG}$ obtained by solving the one-loop RGEs numerically.
Finally, case~(iv) corresponds to the simple recipe
proposed above, in which we evaluate $(\mhl^2)_{\rm 1LL}$
by setting $m_t$ to the running top quark mass at the scale
$\mu_t$. For completeness, we also include
case~(v), where we apply the same recipe to
$(\mhl^2)_{\rm 1LT}$.

The following general features are
noteworthy.  First, we observe that over the region of $\msusy$ shown,
$(\mhl)_{\rm 1RG}\simeq (\mhl)_{\rm 1LL}(m_t(\mu_t))$.
Second, the difference between $(\mhl)_{\rm 1LL}$
and $(\mhl)_{\rm 1RG}$ is non-negligible for even moderate values of
$\msusy$;
neglecting RG-improvement can lead to an overestimate of $\mhl$ which
can be as large as 10 GeV
(for $\msusy>2$~TeV, the deviation grows even larger).  Finally,
note that although the simplest approximation, $(\mhl)_{\rm 1LT}$,
reflects the dominant source of radiative corrections, it yields
the largest overestimate of the light Higgs boson mass.

\section{Additional Complications: Supersymmetric Thresholds}

In the analysis of the previous section, we assumed that all
supersymmetric particle masses were roughly equal and 
substantially larger than $\mz$.
To account for a non-degenerate supersymmetric
spectrum, we must recompute the RGEs in steps starting from
$\mu=\msusy$ and ending at $\mz$. Every
time the threshold of a supersymmetric particle is passed, we integrate
it out of the theory, and determine a new set of RGEs for the new
effective theory.  Eventually, when
we pass below the lightest supersymmetric threshold, we regain the RGEs
of the Standard Model given in \eq{rges}.  We can solve
iteratively for $\lambda(\mz)$ as we did in the previous section, but
now using the more complicated set of RGEs.
Explicit formulae can be found in \refs{llog}{hhh}.

However, the above procedure fails to incorporate the effects of squark
mixing.  Since the most important contribution to the Higgs mass
radiative corrections arises from the incomplete cancelation of the top
quark and top squark loops, it is important to examine
this sector more closely.  First, we define our notation. The
physical top squark squared-masses (in the $v_1=0$ model)
are eigenvalues of the following two $2\times 2$ matrix
\begin{equation}
\left(\begin{array}{cc}
  M_{Q}^2+m_t^2-\mz^2(\half-e_t \sin^2\theta_W) & m_t A_t \\
  m_t A_t & M_{U}^2+m_t^2-\mz^2 e_t \sin^2\theta_W 
\end{array}\right)  
\label{stopmatrix}
\end{equation}
where $e_t=2/3$ and $M_{Q}$, $M_{U}$,
$A_t$ are soft-supersymmetry-breaking parameters.

We shall
treat the squark mixing perturbatively, assuming that the off-diagonal
squark squared-masses are small compared to the diagonal
terms.~\footnote{Formally, we assume that $(M_1^2-M_2^2)/(M_1^2+M_2^2)\ll
1$, where $M_1^2$, $M_2^2$ are the top squark squared-masses.
Thus, we demand that $m_t A_t/\msusy^2\ll 1$.}
The perturbative effect of squark mixing is to modify the
supersymmetric relation between the Higgs quartic coupling and the gauge
couplings [\eq{boundary}].  Such modifications arise from one loop squark
corrections to the Higgs quartic self-coupling via: (i) corrections to
the scalar two-point function on the external legs; (ii) triangle graphs
involving two trilinear Higgs-squark-squark interactions and one quartic
Higgs-Higgs-squark-squark interaction; and (iii) box graphs involving
four trilinear Higgs-squark-squark interactions \cite{infn}.  Then,
\eq{boundary} is modified to:
\beq  \label{newboundary}
\lambda(\msusy)=\quarter(g^2+g^{\prime
2})+\delta\lambda_2+\delta\lambda_3+\delta\lambda_4\,,
\eeq
where the $\delta\lambda_i$ arise from the three sources quoted above.
Explicitly,
\beqa
\delta\lambda_2 &=& {-3(g^2+g^{\prime 2})\over 32\pi^2}A_t^2 h_t^2
B(M_Q^2,M_U^2)\,,\nonumber \\
\delta\lambda_3 &=& {3\over 32\pi^2}\Bigl[4h_t^4 A_t^2 h(M_Q^2,M_U^2)
+g^2 h_t^2 A_t^2 p_t(M_Q^2,M_U^2)\Bigr]\,, \nonumber \\
\delta\lambda_4 &=& {3\over 16\pi^2}h_t^4 A_t^4 g(M_Q^2,M_U^2)\,,
\eeqa
where
\beqa \label{functiondefs}
B(a,b)&\equiv &{1\over (a-b)^2}\left[\half\left(a+b\right)-{ab\over
a-b}\ln\left({a\over b}\right)\right]\,,\nonumber \\
h(a,b)&\equiv &{1\over a-b}\ln\left({a\over b}\right)\,,\nonumber \\
f(a,b)&\equiv & {-1\over (a-b)}\left[1-{b\over a-b}\ln\left({a\over b}
\right)\right]\,, \nonumber \\
g(a,b)&\equiv & {1\over (a-b)^2}\left[2-{a+b\over a-b}\ln\left({a\over b}
\right)\right]\,, \nonumber \\
p_t(a,b)&\equiv &f(a,b)+2e_t \sin^2\theta_W (a-b)g(a,b)\,.
\eeqa

For simplicity, consider the case of $M_Q=M_U\equiv\msusy$.
Using $B(a,a)= 1/6a$, $h(a,a)=1/a$, $f(a,a)=-1/2a$ and $g(a,a)=-1/6a^2$,
\eq{newboundary} becomes:
\beq  \label{mixboundary}
\lambda(\msusy)=\quarter(g^2+g^{\prime
2})+{3h_t^4 A_t^2\over 8\pi^2 \msusy^2}\left[1-{A_t^2\over
12\msusy^2}\right]\,.
\eeq
Note that the correction term due to squark mixing has a maximum when
$A_t=\sqrt{6}\msusy$.  This relation is often called the maximal mixing
condition, since it corresponds to the point at which the one-loop
radiative corrections to $\mhl^2$ are maximal.

Using the new boundary condition, we may repeat the analysis of the
previous section and recompute $\mhl^2$.
At one loop, the effect of the squark mixing is simply additive.  That
is, the modification of $\mhl^2$ due to squark mixing
at one loop is given by:
$(\Delta m_h^2)_{\rm 1mix}=
(\delta\lambda_2+\delta\lambda_3+\delta\lambda_4)v^2$.
At two-loops, we solve for $\lambda(\mz)$ by iterating the RGE for
$\lambda(\mu)$ twice as in the previous section.
However, the boundary condition for $\lambda(\msusy)$ has been
altered, and this modifies the computation.  The end result is
\beq \label{mhmixlog}
(\Delta m_h^2)_{\rm mix}={3g^2m_t^4 A_t^2\over
8\pi^2\mw^2\msusy^2}\left(1-{A_t^2\over 12\msusy^2}\right)
\left[1+2\left(\gamma_v+
        \frac{\beta_{h_t^2}}{h_t^2}\right)\,
        \ln\left(\frac{\msusyy}{m_t^2}\right) \right]
\eeq
{\it i.e.}, $(\Delta m_h^2)_{\rm mix}$ acquires a logarithmically-enhanced
piece at two loops. 
In this approximation, the maximum in $(\Delta m_h^2)_{\rm mix}$
at $A_t=\sqrt{6}\msusy$ is not shifted.
However, this method does {\it not} pick up any 
non-logarithmically-enhanced 
two-loop terms proportional to $A_t$.  To obtain such terms,
one would have to perform a two-loop computation in order to find
the necessary two-loop terms that modify the boundary condition
[\eq{newboundary}].

It is again possible to absorb the two-loop singly-logarithmic term into
the one-loop contribution, $(\Delta m_h^2)_{\rm 1mix}$, by an
appropriate choice of scale for the top-quark mass.  The end result is
quite simple:
\beq \label{deltamix2}
(\Delta m_h^2)_{\rm mix}={3g^2m_t^4(\msusy) A_t^2\over
8\pi^2\mw^2\msusy^2}\left(1-{A_t^2\over 12\msusy^2}\right)\,.
\eeq
That is $(\Delta m_h^2)_{\rm mix}=(\Delta m_h^2)_{\rm
1mix}(m_t(\mu_{\widetilde t}))$, where
the appropriate choice of scale in this case is $\mu_{\widetilde
t}\equiv \msusy$.  The difference from the previous case
[where $\mu_t=\sqrt{m_t\msusy}$] arises due to the extra factor of 2
multiplying the two-loop singly-logarithmic term in \eq{mhmixlog} [compare
this with \eq{lambdatwice}].  Physically, $\mu_{\widetilde t}=\msusy$
corresponds to the scale at which the squarks decouple and
the boundary condition [\eq{mixboundary}] is modified due to squark
mixing.

To illustrate the above results, we compare in Fig.~\ref{hhhfig4}
the value of $\mhl$ as a function of $A_t$
based on the five cases exhibited in Fig.~\ref{hhhfig1}.
Specifically, the effects of $(\Delta m_h^2)_{\rm mix}$ are included at
the one-loop level in cases (i) and (ii), while cases (iv) and (v) make
use of the improved result given by \eq{deltamix2}.
In the full RG-improved result [case (iii)], the RGE for $\lambda(\mu)$
is computed numerically using the modified boundary condition
[\eq{newboundary}].  We see that
$(\mhl^2)_{\rm 1RG}\simeq (\mhl^2)_{\rm
1LL}(m_t(\mu_t))+(\Delta\mhl^2)_{\rm 1mix}(m_t(\widetilde\mu_t))$.
Thus, once again a simple recipe provides an
excellent approximation to the numerically-integrated RG-improved result
over the entire region of the graph.
Note that the maximal value of $\mhl$ occurs for $|A_t|\simeq
2.4\msusy$. The solid or dash-dotted line provides our best mass estimate,
and we conclude that $\mhl\lsim 125$~GeV for $\msusy\leq 1$~TeV.
Similar results were also obtained by Carena {\it et al.} \cite{carena}.


\begin{figure}[htb]
\centerline{\psfig{file=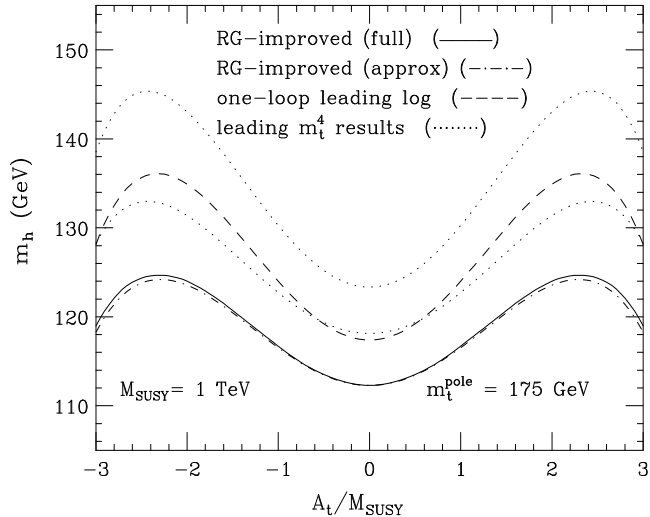,width=8.5cm,height=6.8cm,angle=90}}
\caption{The upper bound to the mass of the light CP-even Higgs boson of
the MSSM plotted as a function of $A_t/\msusy$.  Squark-mixing effects
are incorporated as described in the text. See the caption to
Fig.~\ref{hhhfig1}.}
\label{hhhfig4}
\end{figure}

During the past year, two groups have computed the $A_t$ dependence of
$\mhl$ at the two-loop level.  \Rref{hollik} has performed a
diagrammatic two-loop computation which includes all terms of ${\cal
O}(\alpha_s$), as a function of $\tan\beta$.  \Rref{zhang}
uses an effective potential approach to extend the computation of
\Ref{hempfhoang} and compute directly the two-loop squark mixing
contributions in the $v_1=0$ model.  These results show that the $A_t$
dependence of $\mhl$ is modified slightly at two loops: the maximal
squark mixing point occurs at $A_t\simeq 2\msusy$, a value somewhat
below the result noted above.  Moreover, the value of $\mhl$ at maximal
squark mixing is slightly higher than the one shown in Fig.~\ref{hhhfig4};
for $\msusy=1$~TeV, the maximal value of $\mhl$ is found to be close to
$\mhl\simeq 130$~GeV.
Presumably, these results are due to genuine
two-loop non-logarithmically enhanced terms proportional to a power
of $A_t^2/\msusy^2$.
An important check of the calculations presented in \refs{hollik}{zhang}
would be to explicitly verify the two-loop logarithmically-enhanced 
contribution exhibited in \eq{mhmixlog}.  

\section{Conclusions}
I have described in detail the theoretical basis for the computation of
the upper bound of the mass of the light CP-even Higgs boson of the
MSSM.  It suffices to consider the limiting case of $v_1=0$ which
considerably simplifies the analysis.
I explained how one can use renormalization
group methods to provide a short-cut for obtaining the leading one-loop
and two-loop contributions to $\mhl$.  These methods can also be
generalized to the full MSSM Higgs sector at arbitrary $\tan\beta$.
Further details and references can be found in \Ref{hhh}.

As a result of the work by many groups during this past decade, we
believe that the predicted value of $\mhl$ as a function of the MSSM
parameters is accurately predicted within an uncertainty of a few GeV.
Simple analytic formulae provide an excellent representation of the
known results over a large range of the MSSM parameter
space \cite{carena,hhh}.
The present partially known two-loop information is essential to this
conclusion and provides confidence that there are no surprises lurking
in some corner of the supersymmetric parameter space.  Some clarification is
still needed to understand more completely the dependence on the squark
mixing parameters.


\section*{Acknowledgments}
This paper is based on a collaboration with Ralf Hempfling and Andre
Hoang.  I have learned much from their wisdom.  I would also like to
thank Joan Sola for his kind hospitality during my visit to Barcelona and
RADCOR-98.   This work is partially supported by a grant from the 
U.S. Department of Energy.

\begin{sloppy}
\begin{raggedright}
\def\app#1#2#3{{\sl Act. Phys. Pol. }{\bf B#1} (#2) #3}
\def\apa#1#2#3{{\sl Act. Phys. Austr.}{\bf #1} (#2) #3}
\def\ppnp#1#2#3{{\sl Prog. Part. Nucl. Phys. }{\bf #1} (#2) #3}
\def\npb#1#2#3{{\sl Nucl. Phys. }{\bf B#1} (#2) #3}
\def\jpa#1#2#3{{\sl J. Phys. }{\bf A#1} (#2) #3}
\def\plb#1#2#3{{\sl Phys. Lett. }{\bf B#1} (#2) #3}
\def\prd#1#2#3{{\sl Phys. Rev. }{\bf D#1} (#2) #3}
\def\pR#1#2#3{{\sl Phys. Rev. }{\bf #1} (#2) #3}
\def\prl#1#2#3{{\sl Phys. Rev. Lett. }{\bf #1} (#2) #3}
\def\prc#1#2#3{{\sl Phys. Reports }{\bf #1} (#2) #3}
\def\cpc#1#2#3{{\sl Comp. Phys. Commun. }{\bf #1} (#2) #3}
\def\nim#1#2#3{{\sl Nucl. Inst. Meth. }{\bf #1} (#2) #3}
\def\pr#1#2#3{{\sl Phys. Reports }{\bf #1} (#2) #3}
\def\sovnp#1#2#3{{\sl Sov. J. Nucl. Phys. }{\bf #1} (#2) #3}
\def\jl#1#2#3{{\sl JETP Lett. }{\bf #1} (#2) #3}
\def\jet#1#2#3{{\sl JETP Lett. }{\bf #1} (#2) #3}
\def\zpc#1#2#3{{\sl Z. Phys. }{\bf C#1} (#2) #3}
\def\ptp#1#2#3{{\sl Prog.~Theor.~Phys.~}{\bf #1} (#2) #3}
\def\nca#1#2#3{{\sl Nouvo~Cim.~}{\bf#1A} (#2) #3}
\def\hpa#1#2#3{{\sl Helv.~Phys.~Acta~}{\bf #1} (#2) #3}
\def\aop#1#2#3{{\sl Ann.~of~Phys.~}{\bf #1} (#2) #3}
\def\fP#1#2#3{{\sl Fortschr.~Phys.~}{\bf #1} (#2) #3}

\end{raggedright}
\end{sloppy}
\end{document}